\documentclass[a4paper,fleqn,usenatbib,twocolumn]{mnras}
\usepackage{pdfpages}
\usepackage[T1]{fontenc}
\usepackage{ae,aecompl}
\usepackage{graphicx}
\usepackage{amsmath}
\usepackage{amssymb}
\usepackage{epstopdf}
\usepackage{natbib}

\title[Dark matter and pulsars: I. Ultracompact minihalos]{Investigating dark matter substructure with pulsar timing\\I. Constraints on ultracompact minihalos}

\author[Hamish A. Clark et al.]{
Hamish A. Clark$^{1}$\thanks{E-mail: hamish.clark@sydney.edu.au (HAC)},
Geraint F. Lewis$^{1}$,
Pat Scott$^{2}$
\\
$^{1}$Sydney Institute for Astronomy, School of Physics A28, The University of Sydney, NSW 2006, Australia\\
$^{2}$Department of Physics, Imperial College London, Blackett Laboratory, Prince Consort Road, London SW7 2AZ, UK
}

\date{Accepted XXX. Received YYY; in original form ZZZ}

\pubyear{2015}

\begin{document}
\label{firstpage}
\pagerange{\pageref{firstpage}--\pageref{lastpage}}
\maketitle

\begin{abstract}
Small-scale dark matter structure within the Milky Way is expected to affect pulsar timing. The change in gravitational potential induced by a dark matter halo passing near the line of sight to a pulsar would produce a varying delay in the light travel time of photons from the pulsar. Individual transits produce an effect that would either be too rare or too weak to be detected in 30-year pulsar observations. However, a population of dark matter subhalos would be expected to produce a detectable effect on the measured properties of pulsars if the subhalos constitute a significant fraction of the total halo mass. The effect is to increase the dispersion of measured period derivatives across the pulsar population. By statistical analysis of the ATNF pulsar catalogue, we place an upper limit on this dispersion of $\log \sigma_{\dot{P}} \leq -17.05$. We use this to place strong upper limits on the number density of ultracompact minihalos within the Milky Way. These limits are completely independent of the particle nature of dark matter.
\end{abstract}


%%%%%%%%%%%%%%%%%%%%%%%%%%%%%%%%%%%%%%%%%%%%%%%%%%

%%%%%%%%%%%%%%%%% BODY OF PAPER %%%%%%%%%%%%%%%%%%

\section{Introduction}
Dark matter structure is thought to have been seeded by random density perturbations in the early universe, collapsing well after recombination into the first gravitationally bound dark matter halos. These small scale structures are expected to have hierarchically merged into the larger structures we see today (see e.g.\ \citealt{Lemoine} and \citealt{Mo} on inflationary cosmology and structure formation, respectively). From CMB observations the spectrum of these primordial perturbations is expected to be nearly scale-free; the power is nearly equal on all spatial scales \citep{WMAP2013, Planck2013}. However, current limits only strongly constrain these fluctuations on very large scales (wavenumbers $k \lesssim 3$ Mpc$^{-1}$). On much smaller scales, many cosmological theories beyond the standard model predict that the power spectrum may deviate significantly from the simple Harrison--Zel'dovich (scale-free) model \citep{Adams-Ross, Adams-Cresswell, Ashoorioon, Erickcek-Sigurdson}. 

If significant additional power is present on smaller scales, more fluctuations of very large amplitude ($\delta \gtrsim 0.3$) will be produced than otherwise.  Such large-amplitude fluctuations rapidly collapse to form primordial black holes \citep[PBHs;][]{Carr-Hawking, Carr}. While there are tight constraints on the abundance of these rare objects, it was recently proposed that smaller amplitude fluctuations ($0.3 \gtrsim \delta \gtrsim 10^{-3}$) may give rise to dense dark matter structures known as ultracompact minihalos \citep[UCMHs;][]{Berezinsky2003, Berezinsky12, Berezinsky13, Ricotti, SS09}. Instead of collapsing directly as a black hole, these perturbations grow by gravitational accretion like any other density perturbation until they collapse: logarithmically during radiation domination, and linearly during matter domination. Unlike regular perturbations, the large initial value of the overdensity means that they enter the non-linear regime of growth (i.e.\ collapse) far earlier than do regular $\delta \sim 10^{-5}$ perturbations responsible for most of the large-scale structure seen today.  Due to this early time of collapse, the infall of dark matter onto UCMHs is essentially radial, and their structure will have a very steep density profile as a result \citep[$\rho \propto r^{-9/4}$;][]{Bertschinger,Ricotti}. Although UCMHs continue to accrete both dark matter and baryonic matter after recombination up until the current era, this steep profile means that they are not expected to be tidally disrupted during the course of their evolution \citep{Berezinsky2006,Berezinsky2008} -- to the extent that the probability of survival through to the modern era is essentially unity for all UCMHs that we consider in this paper.

The present-day mass of a UCMH or PBH may be directly mapped to the wavenumber of the primordial fluctuation that originally seeded it, as the wavenumber of a mode re-entering the horizon at any time after inflation depends on the horizon size, which sets the initial mass of the overdense region. Because they require a smaller initial overdensity than PBHs, UCMHs are expected to be produced in far greater numbers than PBHs. UCMHs are therefore a very promising direct link to the conditions of the early universe. By constraining their abundance, we can place limits upon the conditions that would lead to their formation \citep[e.g.][]{JG10, Bringmann, Shandera, Berezinsky2011, Anthonisen}.

Non-detection of PBHs has been used to place very weak limits on primordial curvature perturbations, of $\log{\mathcal{P}_\mathcal{R}} \lesssim -1.5$ for wavenumbers in the range $10^{-2} \lesssim k \lesssim 10^{19}$ Mpc$^{-1}$ \citep{Josan,Carr-Kohri,Alabidi}. Non-detection of UCMHs by gamma-ray searches has been used to constrain their present-day number density, improving this limit significantly, but over a far smaller range of scales: $\log\mathcal{P}_\mathcal{R} \lesssim -6.5$ for wavenumbers in the range $3 \lesssim k \lesssim 10^{7}$ Mpc$^{-1}$ \citep{Bringmann}. It is important to note, however, that the limits from gamma-ray searches depend entirely on the assumption that dark matter can self-annihilate; indeed, UCMHs have been studied extensively as targets for indirect detection of dark matter due to their extremely steep density profiles \citep{SS09, Lacki10, Yang11c, Yang12, Yang11a, Yang13a, Yang13c, Yang13b, Zhang11, Zheng14}.

Gravitational lensing has long been used as an important tool in the detection of dark matter, as gravity is the only force that we are certain dark matter interacts via.  Non-observation of characteristic changes in the position or light curve of a star due to intervening masses (known respectively as astrometric and photometric microlensing) have been used or proposed as means to weakly constrain the abundance of both PBHs \citep{Tisserand, Wyrzykowski2011a,Wyrzykowski2011b, Griest} and UCMHs \citep{Ricotti, Li, Zackrisson12}. However, a confirmed detection of an individual object would require either extreme sensitivity or high abundance, so it is not expected that these methods will be able to effectively constrain the properties of the primordial power spectrum in the near future (unless something like the proposed THEIA satellite mission flies).

As a more sensitive alternative, it has been proposed that substructure could be detected by measuring the effect of an intervening mass on the timing of a millisecond pulsar \citep{Siegel}. This `time-delay lensing' uses the increased travel time of a light ray that passes through a changing gravitational potential, known as the Shapiro effect. Although it is not possible to directly measure the delay due to a static mass, a dark matter halo that moves between an observer and a pulsar would cause the pulse frequency to appear to decrease as the lens travels toward the line of sight, and increase again as the lens moves away. Even though this is a very weak effect, millisecond pulsars can provide extremely accurate clocks -- in many cases, significantly more accurate than atomic clocks -- so it may be expected that large mass subhalos would indeed produce a detectable effect on pulsar timing.

Here, we use time-delay lensing to derive improved limits on the number density of UCMHs within the Milky Way. We first describe the analytical method for calculating the time delay produced by a UCMH (Section \ref{tdels}).  We then compute the probability for a dark matter halo to transit the line of sight to a pulsar, and to observe the event by the delay of the pulsed emission (Section \ref{prob}).  We predict the population impact of this effect to be a type of Gaussian noise present in all pulsar period derivative ($\dot{P}$) measurements, and describe a novel method for using this noise to constrain the properties of dark matter substructure (Section \ref{pdotmethod}).  We use the resulting limits on $\dot{P}$-noise to set limits on the number density of UCMHs within the Milky Way (Section \ref{ucmhnum}), then conclude and summarize (Section \ref{conc}).

\section{Method}
\subsection{Time-Delay Lensing}
\label{tdels}
The Shapiro time delay of a light ray may be investigated by considering the travel time of the light from its source. In the case of an unlensed system, this is simply the proper distance divided by the speed of light. However, in the lensed case, this will differ by
\begin{equation}
\Delta t_{\rm ltt} = \Delta t_{\rm geo} + \Delta t_{\rm pot},
\end{equation}
where $\Delta t_{\rm geo}$ is the geometric component of the delay, i.e. the change in the light ray's path length due to lensing. $\Delta t_{\rm pot}$ is the gravitational potential contribution, which may be calculated from the Newtonian potential through which the light ray passes, as an integral along its path \citep{Petters}
\begin{equation}
\Delta t_{\rm pot} = -\frac{2}{c^3}\int\limits_C\varphi(r)\,\mathrm{d}s.
\end{equation}
Here $c$ is the speed of light, $\varphi$ is the Newtonian potential, $r$ is the radius from the centre of the halo, and $C$ is the path of the light beam, parametrized by $\mathrm{d}s$.

In the case of lensing of sources within the Milky Way by small scale dark matter halos, the deflection of the light ray will be negligible, and so the difference between a lensed system and an unlensed one will be significantly greater in the potential term than the geometric. We may therefore approximate the path of the light ray as a straight line; we refer to this as the zero deflection approximation.
The line integral may then be solved by setting $r = \sqrt{s^2 + b^2}$, so that
\begin{equation}
\Delta t_{\rm ltt} = \Delta t_{\rm pot} = \frac{2}{c^3}\int_{D_{\rm ds}}^{-D_{\rm d}}\varphi(s)\,\mathrm{d}s,
\label{zerodef}
\end{equation}
where we define the position of the lens along the path as $s=0$. Here, $b$ is the impact parameter of the beam, $D_{\rm d}$ is the distance from observer to lens, and $D_{\rm ds}$ is the distance from lens to source.

The Newtonian gravitational potential for any extended spherically symmetric mass with radially varying density is
\begin{equation}
\label{potential}
\varphi(r) = -4\pi G\left[\frac{1}{r}\int_0^r\rho(r')r'^2dr' + \int_r^\infty\rho(r')r'dr'\right].
\end{equation}
Under the approximation of zero deflection, the light travel time from source to observer may be found if the Newtonian potential is easily integrable. Here we are specifically seeking to investigate dark matter substructure that that has not been amenable to detection by standard lensing effects like source magnification or shear, both of which rely on non-negligible light deflection. The assumption of zero deflection is hence not only mathematically convenient in this case, but is implied by the problem itself.

\subsubsection{Ultracompact Minihalos}

UCMHs have been predicted to have density profile \citep{Ricotti2008,Ricotti}
\begin{equation}
\label{ucmhrho}
\rho(r,z) = \frac{\kappa_s(z)}{r^{9/4}},
\end{equation}
where
\begin{equation}
\kappa_s(z) \equiv \frac{3f_\chi M_{h}(z)}{16\pi R_{h}(z)^{\frac{3}{4}}}. 
\end{equation}
Here $f_\chi \equiv \Omega_\chi / \Omega_m$ is the ratio of dark matter to matter in the universe, $M_h$ is the mass of the halo at redshift $z$, and $R_h$ is the radius of the halo, 
\begin{equation}
\frac{R_h (z)}{\textrm{pc}} = 0.019\left(\frac{1000}{z+1}\right)\left(\frac{M_h(z)}{M_\odot}\right)^{1/3}.
\end{equation}

The approximation of purely radial infall breaks down at small $r$, so we take the halo to have a flattened density profile within a core of radius $r_c$ at zero redshift \citep{Bringmann}, given by
\begin{equation}
\frac{r_c}{R_h^0} \approx 2.9 \times 10^{-7} \left(\frac{1000}{z_c+1} \right)^{2.43}\left(\frac{M_h^0}{M_\odot} \right)^{-0.06},
\end{equation}
where $z_c$ is the redshift at which the halo collapsed, $M_h^0$ is the present day mass, and $R_h^0$ is the present day halo radius. In the case of annihilating dark matter, the density is truncated at a slightly lower value, increasing the radius of the core, and decreasing the total mass of the halo by less than a percent -- having a negligible effect on the time delay. Therefore, for the remainder of this paper we consider only the density truncation at $r_c$. The particle nature of dark matter can also impact the possible masses of UCMHs formed in the early Universe, making any limits we draw invalid below a certain cutoff mass.  This mass can vary from $10^{-3}$ to $10^{-11} M_\odot$ for typical WIMPs \citep{Bringmann2009}.

\begin{figure}
\centering
\includegraphics[width=\columnwidth]{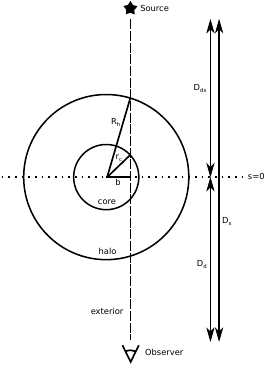}
\caption{Schematic diagram of undeflected lensing by an ultracompact minihalo. The photon path is shown as a dashed line, passing from source, through the different regions of the lens, to observer, with impact parameter $b$. Size of core and halo are exaggerated for visibility.}
\label{lens-schematic}
\end{figure}

From Eqs.\ \ref{potential} and \ref{ucmhrho}, we find the gravitational potential induced by a UCMH to be
\begin{align}
\varphi(r<r_c)     &= \frac{2\pi G\kappa_s}{3}\left(\frac{r^2}{r_c^{9/4}} - \frac{27}{r_c^{1/4}} + \frac{24}{R_h^{1/4}}\right),\\
\varphi(r_c<r<R_h) &= \frac{4\pi G\kappa_s}{3}\left(\frac{12}{R_h^{1/4}} + \frac{3r_c^{3/4}}{r} - \frac{16}{r^{1/4}}\right),\\
\varphi(R_h<r)     &= \frac{4\pi G\kappa_s}{3r}\left(3r_c^{3/4} - 4R_h^{3/4}\right).
\end{align}

By integrating these potentials along the line of sight (Eq.\ \ref{zerodef}) under the assumption of zero deflection, and defining the lens position along the LOS as $s=0$, we find the potential component of the light travel time passing from $s_1$ to $s_2$ within each region of the halo, as shown in Fig.\ \ref{lens-schematic}. 
\vspace{2mm}\\i) Outside of the halo:
\begin{align}
t_{ext}(s_1,s_2) &= \frac{8\pi G\kappa_s}{3c^3}\left(3r_c^{3/4} - 4R_h^{3/4}\right)\ln\left(\frac{\sqrt{b^2 + s_2^2}+s_2}{\sqrt{b^2 + s_1^2}+s_1}\right),
\end{align}
\\ii) Within the halo:
\begin{align}
t_{halo}(s_1,s_2) &= \frac{8\pi G\kappa_s}{3c^3}\Bigg[\frac{12(s_2-s_1)}{R_h^{1/4}} - \frac{16\left[s_2F(s_2) - s_1F(s_1)\right]}{b^{1/4}}\nonumber\\&+3r_c^{3/4}\ln\left(\frac{\sqrt{b^2 + s_2^2}+s_2}{\sqrt{b^2 + s_1^2}+s_1}\right)\Bigg],
\end{align}
where $F(s) \equiv {}_2F_1\left(\frac{1}{8},\frac{1}{2};\frac{3}{2};-\frac{s^2}{b^2}\right)$ is a Gaussian hypergeometric function.
\vspace{2mm}\\iii) Within the core:
\begin{align}
t_{core}(s_1,s_2) &= \frac{4\pi G\kappa_s}{c^3}(s_2-s_1)\Bigg[\frac{8}{R_h^{1/4}} - \frac{9}{r_c^{1/4}}\nonumber \\&+ \frac{(3b^2 + s_1^2 + s_1s_2 + s_2^2)}{9r_c^{9/4}}\Bigg].
\end{align}

The total delay may then be found as the sum of each section through which the light ray passes. For example, the total potential delay of a light ray originating outside of the halo, passing through it (without intersecting the core), and received by an observer on the exterior of the halo may be found by:
\begin{align}
t_{\rm pot} &= t_{ext}\left(D_{\rm ds}, \sqrt{R_h^2 - b^2}\right) \nonumber\\&+ t_{halo}\left(\sqrt{R_h^2 - b^2}, -\sqrt{R_h^2 - b^2}\right) \nonumber\\&+ t_{ext}\left(-\sqrt{R_h^2 - b^2}, -D_{\rm d}\right).
\end{align}

\subsubsection{Navarro-Frenk-White Subhalos}

Similarly, we can also calculate the light travel time for the commonly considered Navarro-Frenk-White (NFW) dark matter density profile. NFW halos have density profile
\begin{equation}
\rho(r) = \frac{\delta_c\rho_c}{r/r_s\left(1+r/r_s\right)^2},
\end{equation}
where 
\begin{equation}
\delta_c  = \frac{200}{3}\frac{c^3}{\ln(1+c)-c/(1+c)},
\end{equation}
$\rho_c = 3H_0^2 / 8\pi G$ is the critical density for closure of the universe at redshift $z = 0$, $H_0$ is the present value of the Hubble constant, $G$ is the gravitational constant and $r_s = r_{200}/c$ is the scale radius.  Here $c$ is the concentration parameter and $r_{200}$ is the virial radius, taken as the radius beyond which the halo is truncated as $\rho(r>r_{200}) = 0$.

We adopt the fitting function for the mass-concentration relation for NFW halos in the low-redshift regime given by \citet{Correa}, allowing each halo to be described solely by its mass, as
\begin{equation}
r_s c\left(M_h\right) = r_{200} = \left(\frac{3M_h}{800\pi \rho_c}\right)^{1/3}.
\end{equation}
The Newtonian gravitational potential of such a halo is
\begin{align}
\varphi(r<r_{200}) &= -4\pi G \delta_c \rho_c r_s^2 \left[\frac{r_s}{r}\ln\left(\frac{r+r_s}{r_s}\right) - \frac{1}{1+c}\right],\\
\varphi(r_{200}<r) &= -\frac{4\pi G \delta_c \rho_c r_s^3}{r}\left(\ln(1+c) - \frac{c}{1+c}\right).
\end{align}
This potential is not analytically integrable, so we integrate it numerically in order to compute the potential time delay.

\subsection{Probability of Individual Subhalo Detection}
\label{prob}
\begin{figure}
\centering
\includegraphics[width=\columnwidth]{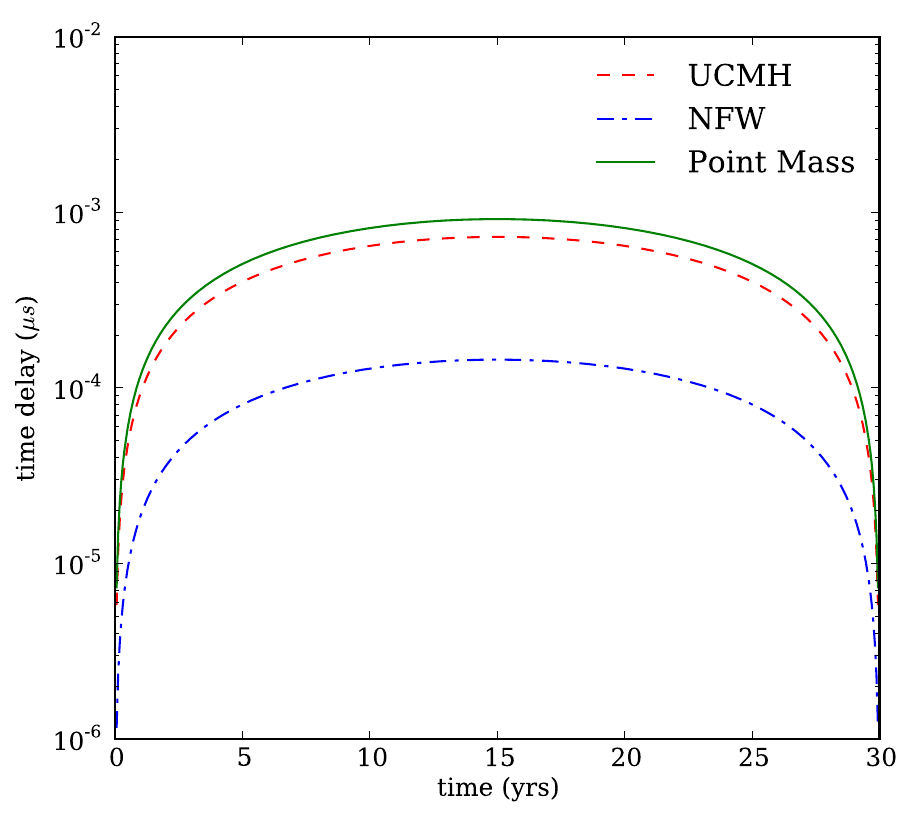}
\caption{An example of a time-delay signal from halos with both UCMH and NFW profiles, as they would be seen in a pulsar's timing residual over a typical pulsar observation time of 30 years, selected such that it passes its point of closest approach to the line of sight after exactly 15 years. Here we have chosen a halo mass $M_h = 1\times10^3\,M_\odot$, impact parameter $b = 10$\,pc, a halo velocity perpendicular to the line of sight $v_{\bot}=200$ km\,s$^{-1}$, observer-pulsar distance $D_s = 1$\,kpc, observer-lens distance $D_{\rm d} = 0.5$\,kpc, and for the NFW profile, concentration parameter $c = 16$.}
\label{tdel}
\end{figure}

Although it is not possible to directly measure a light ray's lensed travel time from a source, changes in the lensing system will cause a change in the light travel time, which in turn will be measured as a change in the residual of the source pulsar's timing. Following the method outlined in section \ref{tdels}, we may iteratively calculate the time delay as a lens moves across the line of sight. We give an example of this signal for both NFW and UCMH dark matter profiles, compared to that due to a point mass, in Fig.\ \ref{tdel}. Due to its significantly steeper gravitational potential, a dark matter subhalo with a UCMH profile would consistently produce a stronger time-delay signal than one following an NFW profile. Due to their steep density profiles, UCMHs are -- for time delay purposes -- essentially pointlike, particularly at masses $\lesssim 1 ~M_\odot$. This makes them prime candidates for any gravitational search for dark matter structure (or alternatively, provide strong opportunities to limit their properties should none be detected).

The rate at which detectable transit events occur for a given pulsar is entirely dependent upon both the number density of halos of a given mass, as well as their velocity distribution. This rate may be predicted by construction of a simple simulation, in which halos are stochastically distributed along the line of sight. Here, we approximate the local dark matter density to be given by a global NFW profile at $r=8$ kpc, with $M_{vir} = 9.4\times 10^{11} M_\odot$, $c=18$, providing a local dark matter density of $0.285 \textrm{ GeV.cm}^{-3}$. We take a fraction $f$ of the local density to be confined within halos of mass $M_h$. We assign each of halo a speed of 200 km\,s$^{-1}$ in the Galactic rest frame, and distribute halos' directions of motion isotropically.  We likewise assume that both the observer and a nearby pulsar (which we set at a distance of 2\,kpc) co-rotate the galactic centre at 220 km\,s$^{-1}$, neglecting the motion of the Earth around the Sun. Allowing each object to continue along its trajectory for 30 years, we record the number of halos that pass within a given distance $r_{\rm max}$ of the line of sight to a pulsar, thereby calculating the approximate rate as a function of $r_{\rm max}$ for a given fraction $f$ and halo mass $M_h$.

As each of these transit events occur independently of one another, their occurrence can be modelled as a homogeneous Poisson process. The probability that at least one event will occur in a pulsar's signal, within a given $r_{\rm max}$ and observation time $\tau$, is
\begin{equation}
P_{\geq 1\ {\rm transit}}(\tau,r_{\rm max}) = 1 - e^{-\tau\lambda(r_{\rm max})},
\label{poisson}
\end{equation}
where $\lambda$ is the event rate determined by simulation. We demand that such an event occur at least once within the 30-year observation data of the 315 pulsars present in the ATNF catalogue \citep{ATNF} at approximately 2\,kpc from Earth (at $\geq$95\% CL). These requirements correspond to a minimum rate of $\lambda \geq 1.0\times10^{-4}$\,yr$^{-1}$.

\begin{figure}
\centering
\includegraphics[width=\columnwidth]{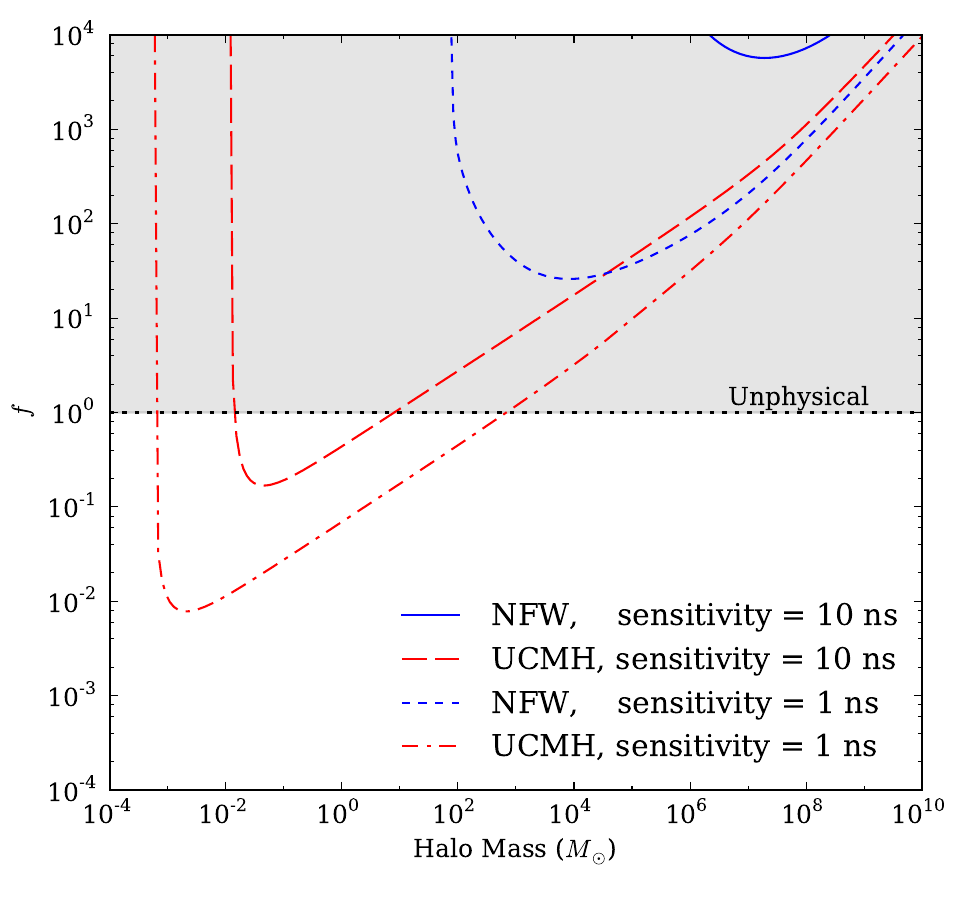}
\caption{The lower limits on the halo mass distribution for which $\ge1$ detectable time-delay lensing event would be present in the timing signal of the 315 pulsars at distance $\sim2$\,kpc in the ATNF catalogue \citep{ATNF}, at 95\% CL.  We show limits for both NFW and UCMH halo profiles, and for 30-year signal amplitude sensitivities of $1\times10^{-8}$\,s and $1\times10^{-9}$\,s. We adopted the local dark matter density throughout the 2\,kpc sphere around the Sun, given by the value at $r=8$\,kpc in a Milky Way NFW profile with $c = 18$ and $M_{vir} = 9.4\times 10^{11} M_\odot$, again corresponding to a local dark matter density of 0.285 GeV/cc \citep{Battaglia2,Battaglia1}. The region in which $f> 1$ represents scenarios that are unphysical, i.e. where the integrated mass of all subhalos would exceed the known mass of the Milky Way.}
\label{halo-detection}
\end{figure}

Following the method we describe in Section \ref{tdels} and taking the optimal case (in which a halo transits along a path perpendicular to the line of sight), we can compute the 30-year amplitude expected from a transit at radius $r_{\rm max}$. Additionally, we simplistically assume that if the 30-year signal amplitude exceeds some sensitivity threshold, the signal may then be taken to be detectable within the pulsar data. Using Brent's root-finding algorithm \citep{Brent}, we vary $f$ for each $M_h$ to find the required event rate, while still producing a signal that exceeds the sensitivity threshold. We show the substructure properties that would be expected to produce at least one detectable event in Fig.\ \ref{halo-detection}, for two different assumed timing sensitivities (1\,ns and 10\,ns).

For all NFW masses that we investigated, the halo fraction that would produce detectable events remains entirely outside of physically reasonable scenarios (i.e.\ $f > 1$).  UCMHs of mass $\approx 10^{-1} M_\odot$ would be expected to be seen in pulsar timing data if they constitute $\gtrsim 5\%$ of the local dark matter. It should be noted, however, that observation of nearby pulsars would be expected to only provide limits on \textit{local} dark matter substructure properties. For NFW halos, this is only indicative of the local clumpiness -- however, UCMHs are immune to tidal disruption and so their number density is expected to follow the global NFW profile of the Milky Way, and thus the local value of $f$ is expected to be the same as the global $f_{\rm MW}$. 

While for some pulsars the time of arrival of a pulse may known with $< 50$ ns precision \citep{HobbsAccuracy}, others achieve far less ($\sim1 \mu$s). Our assumed sensitivity threshold is therefore very optimistic -- with present pulsar data, a detection of any dark matter halo would not be expected. In the future, if a complete search through timing observations of nearby pulsars were to be undertaken at our assumed improved sensitivity, and a timing event found, the discovered object could be identified as an ultracompact minihalo, a compact object (baryonic or otherwise), or as a yet undescribed dark matter structure. Even in the most optimistic case, in which the entire galactic halo is taken to consist entirely of subhalos of equal mass ($f=1$), we find that if a time delay event were to be identified within 30 year data, such an object would not be an NFW halo (at $\geq$ 99.98\% CL), unless timing accuracy significantly exceeded 1 ns. The positive identification of any dark matter substructure, while highly difficult, would have significant implications for our understanding of the conditions present in the early universe, as well as the identity and structure formation of dark matter.

\subsection{Effect on Observed Pulsar Period Derivatives}
\label{pdotmethod}
Searching through pulsar timing data for individual halo transit signals would be an intensive process, and, if a potential detection were found, it would be extremely difficult to differentiate from effects like timing noise, glitches and polynomial fitting errors in periods and period derivatives.  A candidate detection could only be confirmed if the signal were seen in two or more pulsars near to each other on the sky. As we showed in the previous Section, the alignment required to produce a transit effect within the 30 year data is extremely rare anyway, unless the number density of halos is very high.

\begin{figure}
\centering
\includegraphics[width=\columnwidth]{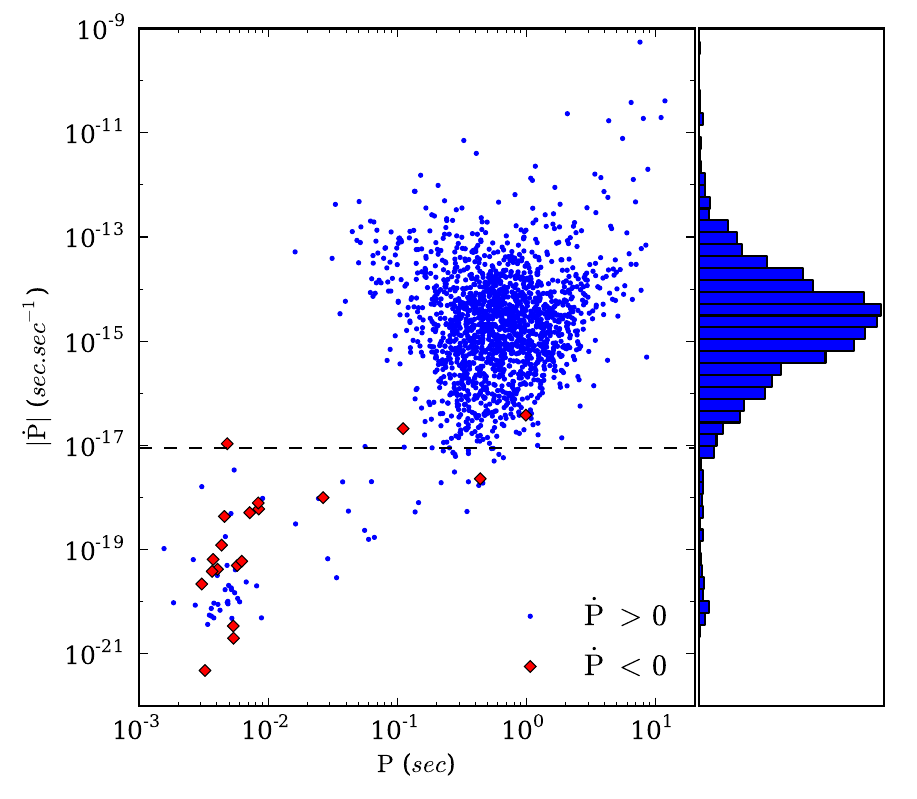}
\caption{The period-period derivative ($\mathrm{P}$-$\dot{\mathrm{P}}$) distribution of non-binary pulsars within the Milky Way from the ATNF pulsar catalogue \citep{ATNF}, co-plotted with a histogram of period derivatives. The upper limit on the standard deviation of the Gaussian $\dot{P}$-noise due to intervening dark matter is displayed as the dashed black line. Binary pulsars have been excluded in order to eliminate potential additional contributions to the period derivative.}
\label{ATNF}
\end{figure}

Alternatively, instead of looking for individual subhalos, searching for net effects on observations of the entire pulsar population due to substructure could be a more effective way of constraining the clumpiness of the Milky Way's dark matter halo. For every halo that completes a transit across the line of sight within 30 years, there are a multitude more that are simply travelling toward or away from their point of closest approach, and thus contribute to the net signal linearly.  That is, it is significantly more likely that the total observed timing effect will be a sum of linear effects, rather than follow the characteristic peaked shape of an individual transit shown in Fig.\ \ref{tdel}. This linear term would ultimately be measured as a contribution to the observed period derivative of a pulsar $\dot{P}_{obs}$,
\begin{equation}
\dot{P}_{obs} \approx \dot{P}_{pulsar} + \sum^N_{i=1}\left(\frac{t_{{\rm pot},i}(t + \Delta t) - t_{{\rm pot},i}(t)}{\Delta t}\right),
\label{pdot}
\end{equation}
for $N$ subhalos, calculated over a time interval $\Delta t$.

The second term of Eq.\ \ref{pdot}, which we refer to as the `$\dot{P}$-noise', is the sum of a function of random variables.  By the central limit theorem, it would therefore present as a Gaussian with mean of zero in the case of large $N$, and variance dependent upon the underlying variables: halo velocity, position, mass, and number density. If the standard deviation of the $\dot{P}$-noise is comparable to the value of the actual period derivative, this additional term would typically dominate over the `true' period derivative of the pulsar.

By examining the distribution of observed pulsar period derivatives from the ATNF catalogue \citep{ATNF}, shown in Fig.\ \ref{ATNF}, it can be seen that the majority of non-binary pulsars are observed to have positive period derivatives.  These appear to follow a log-normal distribution. It is therefore possible to determine how large the standard deviation of the Gaussian $\dot{P}$-noise may be, such that the observed distribution can still be reproduced. For example, if $\sigma_{\dot{P}} \geq  10^{-12}$, the noise would dominate the `true' distribution of pulsars, such that they would be observed as being strongly normal, rather than log-normal.

In order to investigate the upper limit on $\sigma_{\dot{P}}$, we model the `true' distribution of pulsar period derivatives as a log-normal distribution. Given that the luminosity distribution of isolated pulsars has been shown to be log-normal \citep{Faucher, Ridley}, this is a reasonable assumption. This relationship can be understood by pulsars only losing rotational energy at a rate linearly proportional to their luminosity. 

When observed, the true $\dot{P}$ distribution will appear convolved with both the instrumental noise and the time-delay noise,
\begin{equation}
f_{obs}(\dot{P}) = \exp\left[N(\xi, \sigma_{\rm true}^2)\right] \ast N(0, \sigma_{\dot{P}}^2) \ast N(0,\sigma_{\rm I}^2)
\end{equation}
where $N(\mu, \sigma^2)$ is the normal distribution, $\sigma_{\dot{P}}^2$ is the induced variance due to time delay lensing, $\sigma_{\rm I}^2$ is the variance due to instrumental noise, and $\xi$, $\sigma_{\rm true}^2$ are respectively the mean and variance of the `true' log-normal distribution of pulsar period derivatives. This convolution results in the likelihood function for $N$ pulsars of \citep{Hawkins}
\begin{align}
\mathcal{L}(\dot{P}_{\rm obs}|&\sigma_{\dot{P}},\xi,\sigma_{\rm true}) = \prod^N_{i=1}\int_{0}^\infty \frac{1}{2\pi \sigma_{\rm true}y \sqrt{\sigma_{\dot{P}}^2+\sigma_{\rm I, i}^2}}\nonumber\\&\times\exp \bigg[ -\frac{(\ln y - \xi)^2}{2\sigma_{\rm true}^2}-\frac{(\dot{P}_{\textrm {obs}, i}-y)^2}{2(\sigma_{\dot{P}}^2 + \sigma_{\rm I, i}^2)} \bigg]dy.
\label{pdotmodel}
\end{align}

To determine the allowed values of $\sigma_{\dot{P}}$, we fitted the observed values of $\dot{P}_{obs}$ from the ATNF catalogue using Eq.\ \ref{pdotmodel}. Using the publicly-available nested sampling algorithm \textsc{MultiNest} v3.9 \citep{Multinest2,Multinest1,Multinest3}, we scanned over $\log\sigma_{\dot{P}}$, $\xi$ and $\sigma_{\rm true}^2$ with uniform priors. We adopted the values of $\sigma_{\rm I}^2$ given for each individual pulsar by the ATNF catalogue.  

We found the posterior mean of the $\dot{P}$-noise standard deviation, which we obtained by marginalising over the log-normal parameters $\xi$ and $\sigma_{\rm true}^2$, as $\log\sigma_{\dot{P}} = -17.21$ -- with 95\% credible interval upper limit of $-$17.05, and lower limit of $-$17.31. However, millisecond pulsars are not so simply described, and may exhibit a range of complex timing phenomena contributing to their final measured signal.  It therefore should not be assumed that their spread is solely due to intervening dark matter and instrumental error.  As such, we adopt the upper limit implied by the 95\% confidence Bayesian credible interval on $\log \sigma_{\dot{P}}$.  Our limit, $\log\sigma_{\dot{P}} \leq -17.05$, is shown as the black dashed line in Fig.\ \ref{ATNF}. Should the dark matter distribution be such that the $\dot{P}$-noise is equal to this value, it would be expected that approximately half of the pulsars with period derivative beneath this will be observed to have a negative period derivative.

\section{Limits on Ultracompact Minihalo Number Density}
\label{ucmhnum}
We can compare the magnitude of $\dot{P}$-noise expected from a population of UCMHs with a given mass to the upper limit that we found in Section \ref{pdotmethod}, using the method outlined in Section \ref{tdels} and Eq.\ \ref{pdot}.  To this end, we simulated UCMHs distributed stochastically according to a Milky Way-like NFW profile, again adopting $c = 18$, $M_{vir} = 9.4\times10^{11} M_\odot$ from \citet{Battaglia2,Battaglia1}, in order to produce a simulated period derivative contribution.

Galactic dark matter subhalo velocities are currently poorly understood, and so we took a simplified model in which all subhalos have isotropic velocities of magnitude 200\,km\,s$^{-1}$, independent of galactocentric radius. We placed the observer at a galactocentric radius of 8\,kpc from the centre of the Galactic profile, and simulated observations of mock lines of sight to all 1810 non-binary pulsars with known period derivatives in the ATNF catalogue.

As the time delay due to each individual subhalo is additive and independent, it is expected that the contributions of lenses of different masses would likewise be additive. In this way, we constrain the fraction of the Milky Way in UCMHs of each mass independently.  It should therefore be noted that models predicting a \textit{range} of UCMH masses must have their predictions integrated over our single-mass limits in order to properly assess their validity.  

\begin{figure}
\centering
\includegraphics[width=\columnwidth]{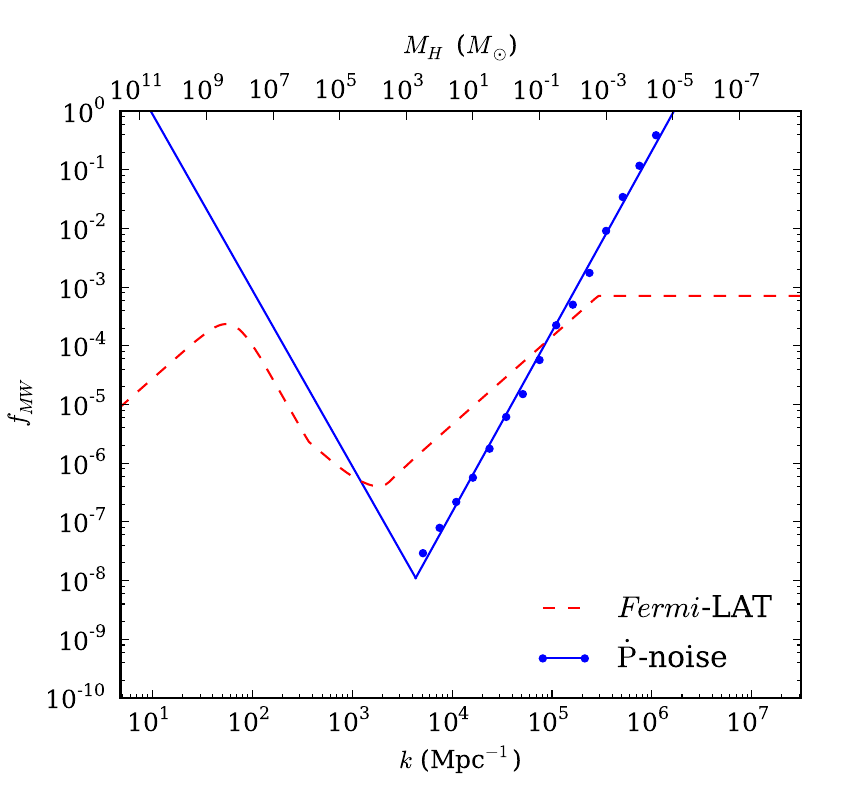}
\caption{The 95\% credible interval upper limits on the fraction of dark matter in the Milky Way contained in UCMHs.  Limits are shown from both pulsar period derivatives and non-detection in gamma-rays by $Fermi$-LAT searches for dark matter annihilation. We map UCMH masses to wavenumber $k$ at horizon entry following the method outlined in \citet{Bringmann}.}
\label{f-lims}
\end{figure}

Varying the fraction of dark matter contained in UCMHs of a given mass, and matching the output $\dot{P}$-noise to our upper limit from pulsar data, we obtain an upper limit on UCMH number density within the Milky Way as a function of mass.  We show this limit in Fig.\ \ref{f-lims}, comparing to those from gamma-ray searches by \citet{Bringmann}, which necessarily assume a specific model for annihilating  dark matter. Mirroring their procedure, we reduce our limits at large masses ($\gtrsim 10^3 M_\odot$), corresponding to case in which the $\dot{P}$-noise does exceed the observational upper limit, but the large-N condition of the central limit theorem no longer holds. By performing a chi-squared test for normality on the time delay distribution at low N, we find that that for $N>12$ the distribution remains normal ($p$-value $<0.05$) for all halo masses that we consider. As a conservative measure, we therefore reduce our limit to $N\geq20$ halos within the Milky Way, as can be seen in Fig.\ \ref{f-lims}. 

Our final limits on UCMH number density are of similar magnitude to those found by gamma-ray searches, with strongest constraint $f \leq 1.1\times 10^{-8}$ at a mass of approximately $1\times10^3 M_\odot$. In fact, our constraints from pulsar timing are significantly stronger throughout the range $10^{-2} \lesssim M_h/M_\odot \lesssim 10^4$ than those found previously. Not only are these the strongest limits on the abundance of UCMH number density to date, they have the additional benefit of being the only strong constraint that does not rely on the assumption that dark matter undergoes annihilation.

Baryonic matter along the line of sight may be expected to produce the same gravitational effects. While additional structure (baryonic or otherwise) would boost this signal beyond that provided by UCMHs alone, we have provided only upper limits on the \textit{total} Gaussian noise, and so our constraint on UCMH abundance still holds true. In addition to gravitational effects, baryonic matter is known to change the speed of light propagation -- providing an additional (potentially varying) delay. The strength of this effect changes on far shorter timescales than that due to dark matter structure, and so does not change the measured period derivative of the pulsar \citep{You2007}. While some pulsars do indeed appear to have a constantly increasing or decreasing dispersion measure, this would have the same effect on the period derivative distribution (at a single wavelength) as the effect we investigate -- adding to the total Gaussian noise.

\section{Conclusions}
\label{conc}
The difference in photon travel time due to an intervening source of gravitational potential has long been known, most famously as the Shapiro delay. It was recently proposed that movement of dark matter structures along the line of sight to a pulsar could potentially produce a measurable change in the pulsar's timing signal. If measurement of these induced shifts were possible, the properties of the intervening structures could potentially be investigated, providing insight into the nature of dark matter and the formation of its structure.

UCMHs are one such predicted form of dark matter substructure, expected to be seeded in the very early universe. As a consequence of their early formation, they have extremely steep density profiles, and consequently steep gravitational potentials, allowing them to provide a strong time-delay signal. UCMHs have been shown to be highly immune to disruption by tidal forces, and so are expected to have persisted from their collapse shortly after matter-radiation equality through to the present day. They therefore provide a unique probe of the conditions of the early universe.

Here we have calculated the probability of detecting an individual halo by pulsar timing.  We find that detections are likely impossible for NFW minihalos, but UCMHs may produce a detectable signal, should the number density of lenses be appropriately high ($f\gtrsim0.01$) for UCMHs in the mass range $10^{-3} \lesssim M_h/M_\odot \lesssim 10^{3}$.

More excitingly, we predict an additive time-delay `noise' on pulsar period derivatives due to a population of dark matter halos. By determining an observational upper limit of $\log\sigma_{\dot{P}} \leq -17.05$ on the observed amount of this noise, we have placed upper limits on the number densities of UCMHs at a range of masses.  While the previous strongest limits rely on the assumption that dark matter can annihilate, the limits we find here are placed by gravitational methods only, and are therefore equally applicable to \textit{any dark matter model}.  Our best limit of $f_{\rm MW} \lesssim 1.1\times10^{-8}$ is more than an order of magnitude better than previous ones.  For masses $10^{-2} \lesssim M_h/M_\odot \lesssim 10^4$, our limits are the strongest available, even compared to earlier model-dependent ones.

While different lensing methodologies appear to have been exhaustively used for investigating the small-scale structure of dark matter, time-delay methods have been mostly overlooked. This new methodology shows that time delays may yet provide the most sensitive measures of lensing to date. We have shown that, although small in amplitude, the gravitational time-delay signal due to UCMHs may indeed yet be seen in pulsar timing observations.  Should a detection be made, these may be used to constrain models of early universe cosmology such as inflation and cosmic strings.

\section*{Acknowledgements}

HAC would like to acknowledge the Australian Postgraduate Awards (APA), through which this work was financially supported. GFL gratefully acknowledges the Australian Research Council (ARC) for support through DP130100117. PS is supported by STFC through the Ernest Rutherford Fellowships scheme.

%%%%%%%%%%%%%%%%%%%%%%%%%%%%%%%%%%%%%%%%%%%%%%%%%%


\bibliography{UCMHPaper1.bib}{}

\begin{thebibliography}{}
\makeatletter
\relax
\def\mn@urlcharsother{\let\do\@makeother \do\$\do\&\do\#\do\^\do\_\do\%\do\~}
\def\mn@doi{\begingroup\mn@urlcharsother \@ifnextchar [ {\mn@doi@}
  {\mn@doi@[]}}
\def\mn@doi@[#1]#2{\def\@tempa{#1}\ifx\@tempa\@empty \href
  {http://dx.doi.org/#2} {doi:#2}\else \href {http://dx.doi.org/#2} {#1}\fi
  \endgroup}
\def\mn@eprint#1#2{\mn@eprint@#1:#2::\@nil}
\def\mn@eprint@arXiv#1{\href {http://arxiv.org/abs/#1} {{\tt arXiv:#1}}}
\def\mn@eprint@dblp#1{\href {http://dblp.uni-trier.de/rec/bibtex/#1.xml}
  {dblp:#1}}
\def\mn@eprint@#1:#2:#3:#4\@nil{\def\@tempa {#1}\def\@tempb {#2}\def\@tempc
  {#3}\ifx \@tempc \@empty \let \@tempc \@tempb \let \@tempb \@tempa \fi \ifx
  \@tempb \@empty \def\@tempb {arXiv}\fi \@ifundefined
  {mn@eprint@\@tempb}{\@tempb:\@tempc}{\expandafter \expandafter \csname
  mn@eprint@\@tempb\endcsname \expandafter{\@tempc}}}

\bibitem[\protect\citeauthoryear{{Adams}, {Ross}  \& {Sarkar}}{{Adams}
  et~al.}{1997}]{Adams-Ross}
{Adams} J.~A.,  {Ross} G.~G.,   {Sarkar} S.,  1997, \mn@doi [Nucl. Phys. B]
  {10.1016/S0550-3213(97)00431-8}, \href
  {http://adsabs.harvard.edu/abs/1997NuPhB.503..405A} {503, 405}

\bibitem[\protect\citeauthoryear{{Adams}, {Cresswell}  \& {Easther}}{{Adams}
  et~al.}{2001}]{Adams-Cresswell}
{Adams} J.,  {Cresswell} B.,   {Easther} R.,  2001, \mn@doi [\prd]
  {10.1103/PhysRevD.64.123514}, \href
  {http://adsabs.harvard.edu/abs/2001PhRvD..64l3514A} {64, 123514}

\bibitem[\protect\citeauthoryear{{Alabidi}, {Kohri}, {Sasaki}  \&
  {Sendouda}}{{Alabidi} et~al.}{2012}]{Alabidi}
{Alabidi} L.,  {Kohri} K.,  {Sasaki} M.,   {Sendouda} Y.,  2012, \mn@doi
  [\jcap] {10.1088/1475-7516/2012/09/017}, \href
  {http://adsabs.harvard.edu/abs/2012JCAP...09..017A} {9, 17}

\bibitem[\protect\citeauthoryear{{Anthonisen}, {Brandenberger}  \&
  {Scott}}{{Anthonisen} et~al.}{2015}]{Anthonisen}
{Anthonisen} M.,  {Brandenberger} R.,   {Scott} P.,  2015, \prd, \href
  {http://adsabs.harvard.edu/abs/2015arXiv150401410A} {92, 023521}

\bibitem[\protect\citeauthoryear{{Ashoorioon}, {Krause}  \&
  {Turzynski}}{{Ashoorioon} et~al.}{2009}]{Ashoorioon}
{Ashoorioon} A.,  {Krause} A.,   {Turzynski} K.,  2009, \mn@doi [\jcap]
  {10.1088/1475-7516/2009/02/014}, \href
  {http://adsabs.harvard.edu/abs/2009JCAP...02..014A} {2, 14}

\bibitem[\protect\citeauthoryear{{Battaglia} et~al.,}{{Battaglia}
  et~al.}{2005}]{Battaglia2}
{Battaglia} G.,  et~al., 2005, \mn@doi [\mnras]
  {10.1111/j.1365-2966.2005.09367.x}, \href
  {http://adsabs.harvard.edu/abs/2005MNRAS.364..433B} {364, 433}

\bibitem[\protect\citeauthoryear{{Battaglia} et~al.,}{{Battaglia}
  et~al.}{2006}]{Battaglia1}
{Battaglia} G.,  et~al., 2006, \mn@doi [\mnras]
  {10.1111/j.1365-2966.2006.10688.x}, \href
  {http://adsabs.harvard.edu/abs/2006MNRAS.370.1055B} {370, 1055}

\bibitem[\protect\citeauthoryear{{Berezinsky}, {Dokuchaev}  \&
  {Eroshenko}}{{Berezinsky} et~al.}{2003}]{Berezinsky2003}
{Berezinsky} V.,  {Dokuchaev} V.,   {Eroshenko} Y.,  2003, \mn@doi [\prd]
  {10.1103/PhysRevD.68.103003}, \href
  {http://adsabs.harvard.edu/abs/2003PhRvD..68j3003B} {68, 103003}

\bibitem[\protect\citeauthoryear{{Berezinsky}, {Dokuchaev}  \&
  {Eroshenko}}{{Berezinsky} et~al.}{2006}]{Berezinsky2006}
{Berezinsky} V.,  {Dokuchaev} V.,   {Eroshenko} Y.,  2006, \mn@doi [\prd]
  {10.1103/PhysRevD.73.063504}, \href
  {http://adsabs.harvard.edu/abs/2006PhRvD..73f3504B} {73, 063504}

\bibitem[\protect\citeauthoryear{{Berezinsky}, {Dokuchaev}  \&
  {Eroshenko}}{{Berezinsky} et~al.}{2008}]{Berezinsky2008}
{Berezinsky} V.,  {Dokuchaev} V.,   {Eroshenko} Y.,  2008, \mn@doi [\prd]
  {10.1103/PhysRevD.77.083519}, \href
  {http://adsabs.harvard.edu/abs/2008PhRvD..77h3519B} {77, 083519}

\bibitem[\protect\citeauthoryear{{Berezinsky}, {Dokuchaev}  \&
  {Eroshenko}}{{Berezinsky} et~al.}{2011}]{Berezinsky2011}
{Berezinsky} V.~S.,  {Dokuchaev} V.~I.,   {Eroshenko} Y.~N.,  2011, \mn@doi
  [\jcap] {10.1088/1475-7516/2011/12/007}, \href
  {http://adsabs.harvard.edu/abs/2011JCAP...12..007B} {12, 7}

\bibitem[\protect\citeauthoryear{{Berezinsky}, {Dokuchaev}  \&
  {Eroshenko}}{{Berezinsky} et~al.}{2012}]{Berezinsky12}
{Berezinsky} V.~S.,  {Dokuchaev} V.~I.,   {Eroshenko} Y.~N.,  2012, \mn@doi
  [Gravitation and Cosmology] {10.1134/S0202289312010045}, \href
  {http://adsabs.harvard.edu/abs/2012GrCo...18...57B} {18, 57}

\bibitem[\protect\citeauthoryear{{Berezinsky}, {Dokuchaev}  \&
  {Eroshenko}}{{Berezinsky} et~al.}{2013}]{Berezinsky13}
{Berezinsky} V.~S.,  {Dokuchaev} V.~I.,   {Eroshenko} Y.~N.,  2013, \mn@doi
  [\jcap] {10.1088/1475-7516/2013/11/059}, \href
  {http://adsabs.harvard.edu/abs/2013JCAP...11..059B} {11, 59}

\bibitem[\protect\citeauthoryear{{Bertschinger}}{{Bertschinger}}{1985}]{Bertschinger}
{Bertschinger} E.,  1985, \mn@doi [\apjs] {10.1086/191028}, \href
  {http://adsabs.harvard.edu/abs/1985ApJS...58...39B} {58, 39}

\bibitem[\protect\citeauthoryear{Brent}{Brent}{1973}]{Brent}
Brent R.,  1973, Algorithms for Minimization Without Derivatives.
Dover Books on Mathematics, Dover Publications

\bibitem[\protect\citeauthoryear{{Bringmann}}{{Bringmann}}{2009}]{Bringmann2009}
{Bringmann} T.,  2009, \mn@doi [New J. Phys.] {10.1088/1367-2630/11/10/105027},
  \href {http://adsabs.harvard.edu/abs/2009NJPh...11j5027B} {11, 105027}

\bibitem[\protect\citeauthoryear{{Bringmann}, {Scott}  \& {Akrami}}{{Bringmann}
  et~al.}{2012}]{Bringmann}
{Bringmann} T.,  {Scott} P.,   {Akrami} Y.,  2012, \mn@doi [\prd]
  {10.1103/PhysRevD.85.125027}, \href
  {http://adsabs.harvard.edu/abs/2012PhRvD..85l5027B} {85, 125027}

\bibitem[\protect\citeauthoryear{{Carr}}{{Carr}}{1975}]{Carr}
{Carr} B.~J.,  1975, \mn@doi [\apj] {10.1086/153853}, \href
  {http://adsabs.harvard.edu/abs/1975ApJ...201....1C} {201, 1}

\bibitem[\protect\citeauthoryear{{Carr} \& {Hawking}}{{Carr} \&
  {Hawking}}{1974}]{Carr-Hawking}
{Carr} B.~J.,  {Hawking} S.~W.,  1974, \mnras, \href
  {http://adsabs.harvard.edu/abs/1974MNRAS.168..399C} {168, 399}

\bibitem[\protect\citeauthoryear{{Carr}, {Kohri}, {Sendouda}  \&
  {Yokoyama}}{{Carr} et~al.}{2010}]{Carr-Kohri}
{Carr} B.~J.,  {Kohri} K.,  {Sendouda} Y.,   {Yokoyama} J.,  2010, \mn@doi
  [\prd] {10.1103/PhysRevD.81.104019}, \href
  {http://adsabs.harvard.edu/abs/2010PhRvD..81j4019C} {81, 104019}

\bibitem[\protect\citeauthoryear{{Correa}, {Wyithe}, {Schaye}  \&
  {Duffy}}{{Correa} et~al.}{2015}]{Correa}
{Correa} C.~A.,  {Wyithe} J.~S.~B.,  {Schaye} J.,   {Duffy} A.~R.,  2015,
  \mn@doi [\mnras] {10.1093/mnras/stv1363}, \href
  {http://adsabs.harvard.edu/abs/2015MNRAS.452.1217C} {452, 1217}

\bibitem[\protect\citeauthoryear{{Erickcek} \& {Sigurdson}}{{Erickcek} \&
  {Sigurdson}}{2011}]{Erickcek-Sigurdson}
{Erickcek} A.~L.,  {Sigurdson} K.,  2011, \mn@doi [\prd]
  {10.1103/PhysRevD.84.083503}, \href
  {http://adsabs.harvard.edu/abs/2011PhRvD..84h3503E} {84, 083503}

\bibitem[\protect\citeauthoryear{{Faucher-Gigu{\`e}re} \&
  {Kaspi}}{{Faucher-Gigu{\`e}re} \& {Kaspi}}{2006}]{Faucher}
{Faucher-Gigu{\`e}re} C.-A.,  {Kaspi} V.~M.,  2006, \mn@doi [\apj]
  {10.1086/501516}, \href {http://adsabs.harvard.edu/abs/2006ApJ...643..332F}
  {643, 332}

\bibitem[\protect\citeauthoryear{{Feroz} \& {Hobson}}{{Feroz} \&
  {Hobson}}{2008}]{Multinest2}
{Feroz} F.,  {Hobson} M.~P.,  2008, \mn@doi [\mnras]
  {10.1111/j.1365-2966.2007.12353.x}, \href
  {http://adsabs.harvard.edu/abs/2008MNRAS.384..449F} {384, 449}

\bibitem[\protect\citeauthoryear{{Feroz}, {Hobson}  \& {Bridges}}{{Feroz}
  et~al.}{2009}]{Multinest1}
{Feroz} F.,  {Hobson} M.~P.,   {Bridges} M.,  2009, \mn@doi [\mnras]
  {10.1111/j.1365-2966.2009.14548.x}, \href
  {http://adsabs.harvard.edu/abs/2009MNRAS.398.1601F} {398, 1601}

\bibitem[\protect\citeauthoryear{{Feroz}, {Hobson}, {Cameron}  \&
  {Pettitt}}{{Feroz} et~al.}{2013}]{Multinest3}
{Feroz} F.,  {Hobson} M.~P.,  {Cameron} E.,   {Pettitt} A.~N.,  2013, preprint,
  \href {http://adsabs.harvard.edu/abs/2013arXiv1306.2144F} {} (\mn@eprint
  {arXiv} {1306.2144})

\bibitem[\protect\citeauthoryear{{Griest}, {Cieplak}  \& {Lehner}}{{Griest}
  et~al.}{2013}]{Griest}
{Griest} K.,  {Cieplak} A.~M.,   {Lehner} M.~J.,  2013, \mn@doi [\prl]
  {10.1103/PhysRevLett.111.181302}, \href
  {http://adsabs.harvard.edu/abs/2013PhRvL.111r1302G} {111, 181302}

\bibitem[\protect\citeauthoryear{Hawkins}{Hawkins}{1991}]{Hawkins}
Hawkins D.~M.,  1991, South African Statist. J., 25, 99

\bibitem[\protect\citeauthoryear{{Hinshaw} et~al.,}{{Hinshaw}
  et~al.}{2013}]{WMAP2013}
{Hinshaw} G.,  et~al., 2013, \mn@doi [\apjs] {10.1088/0067-0049/208/2/19},
  \href {http://adsabs.harvard.edu/abs/2013ApJS..208...19H} {208, 19}

\bibitem[\protect\citeauthoryear{{Hobbs}}{{Hobbs}}{2014}]{HobbsAccuracy}
{Hobbs} G.,  2014, in {Capitaine} N.,  ed., Journ{\'e}es 2013 ``Syst{\`e}mes de
  r{\'e}f{\'e}rence spatio-temporels''. pp 115--119

\bibitem[\protect\citeauthoryear{{Josan} \& {Green}}{{Josan} \&
  {Green}}{2010}]{JG10}
{Josan} A.~S.,  {Green} A.~M.,  2010, \mn@doi [\prd]
  {10.1103/PhysRevD.82.083527}, \href
  {http://adsabs.harvard.edu/abs/2010PhRvD..82h3527J} {82, 083527}

\bibitem[\protect\citeauthoryear{{Josan}, {Green}  \& {Malik}}{{Josan}
  et~al.}{2009}]{Josan}
{Josan} A.~S.,  {Green} A.~M.,   {Malik} K.~A.,  2009, \mn@doi [\prd]
  {10.1103/PhysRevD.79.103520}, \href
  {http://adsabs.harvard.edu/abs/2009PhRvD..79j3520J} {79, 103520}

\bibitem[\protect\citeauthoryear{{Lacki} \& {Beacom}}{{Lacki} \&
  {Beacom}}{2010}]{Lacki10}
{Lacki} B.~C.,  {Beacom} J.~F.,  2010, \mn@doi [\apjl]
  {10.1088/2041-8205/720/1/L67}, \href
  {http://adsabs.harvard.edu/abs/2010ApJ...720L..67L} {720, L67}

\bibitem[\protect\citeauthoryear{Lemoine, Martin  \& Peter}{Lemoine
  et~al.}{2008}]{Lemoine}
Lemoine M.,  Martin J.,   Peter P.,  2008, Inflationary Cosmology.
Lecture Notes in Physics, Springer

\bibitem[\protect\citeauthoryear{{Li}, {Erickcek}  \& {Law}}{{Li}
  et~al.}{2012}]{Li}
{Li} F.,  {Erickcek} A.~L.,   {Law} N.~M.,  2012, \mn@doi [\prd]
  {10.1103/PhysRevD.86.043519}, \href
  {http://adsabs.harvard.edu/abs/2012PhRvD..86d3519L} {86, 043519}

\bibitem[\protect\citeauthoryear{{Manchester}, {Hobbs}, {Teoh}  \&
  {Hobbs}}{{Manchester} et~al.}{2005}]{ATNF}
{Manchester} R.~N.,  {Hobbs} G.~B.,  {Teoh} A.,   {Hobbs} M.,  2005, \mn@doi
  [\aj] {10.1086/428488}, \href
  {http://adsabs.harvard.edu/abs/2005AJ....129.1993M} {129, 1993}

\bibitem[\protect\citeauthoryear{Mo, van~den Bosch  \& White}{Mo
  et~al.}{2010}]{Mo}
Mo H.,  van~den Bosch F.,   White S.,  2010, Galaxy Formation and Evolution.
Cambridge University Press

\bibitem[\protect\citeauthoryear{Petters, Levine  \& Wambsganss}{Petters
  et~al.}{2012}]{Petters}
Petters A.,  Levine H.,   Wambsganss J.,  2012, Singularity Theory and
  Gravitational Lensing.
Progress in Mathematical Physics, Birkh{\"a}user Boston

\bibitem[\protect\citeauthoryear{{Planck Collaboration}}{{Planck
  Collaboration}}{2014}]{Planck2013}
{Planck Collaboration} 2014, \mn@doi [\aap] {10.1051/0004-6361/201321569},
  \href {http://adsabs.harvard.edu/abs/2014A%26A...571A..22P} {571, A22}

\bibitem[\protect\citeauthoryear{{Ricotti} \& {Gould}}{{Ricotti} \&
  {Gould}}{2009}]{Ricotti}
{Ricotti} M.,  {Gould} A.,  2009, \mn@doi [\apj] {10.1088/0004-637X/707/2/979},
  \href {http://adsabs.harvard.edu/abs/2009ApJ...707..979R} {707, 979}

\bibitem[\protect\citeauthoryear{{Ricotti}, {Ostriker}  \& {Mack}}{{Ricotti}
  et~al.}{2008}]{Ricotti2008}
{Ricotti} M.,  {Ostriker} J.~P.,   {Mack} K.~J.,  2008, \mn@doi [\apj]
  {10.1086/587831}, \href {http://adsabs.harvard.edu/abs/2008ApJ...680..829R}
  {680, 829}

\bibitem[\protect\citeauthoryear{{Ridley} \& {Lorimer}}{{Ridley} \&
  {Lorimer}}{2010}]{Ridley}
{Ridley} J.~P.,  {Lorimer} D.~R.,  2010, \mn@doi [\mnras]
  {10.1111/j.1365-2966.2010.16342.x}, \href
  {http://adsabs.harvard.edu/abs/2010MNRAS.404.1081R} {404, 1081}

\bibitem[\protect\citeauthoryear{{Scott} \& {Sivertsson}}{{Scott} \&
  {Sivertsson}}{2009}]{SS09}
{Scott} P.,  {Sivertsson} S.,  2009, \mn@doi [\prl]
  {10.1103/PhysRevLett.103.211301}, \href
  {http://adsabs.harvard.edu/abs/2009PhRvL.103u1301S} {103, 211301}

\bibitem[\protect\citeauthoryear{{Shandera}, {Erickcek}, {Scott}  \&
  {Galarza}}{{Shandera} et~al.}{2013}]{Shandera}
{Shandera} S.,  {Erickcek} A.~L.,  {Scott} P.,   {Galarza} J.~Y.,  2013,
  \mn@doi [\prd] {10.1103/PhysRevD.88.103506}, \href
  {http://adsabs.harvard.edu/abs/2013PhRvD..88j3506S} {88, 103506}

\bibitem[\protect\citeauthoryear{{Siegel}, {Hertzberg}  \& {Fry}}{{Siegel}
  et~al.}{2007}]{Siegel}
{Siegel} E.~R.,  {Hertzberg} M.~P.,   {Fry} J.~N.,  2007, \mn@doi [\mnras]
  {10.1111/j.1365-2966.2007.12435.x}, \href
  {http://adsabs.harvard.edu/abs/2007MNRAS.382..879S} {382, 879}

\bibitem[\protect\citeauthoryear{{Tisserand} et~al.,}{{Tisserand}
  et~al.}{2007}]{Tisserand}
{Tisserand} P.,  et~al., 2007, \mn@doi [\aap] {10.1051/0004-6361:20066017},
  \href {http://adsabs.harvard.edu/abs/2007A%26A...469..387T} {469, 387}

\bibitem[\protect\citeauthoryear{{Wyrzykowski} et~al.,}{{Wyrzykowski}
  et~al.}{2011a}]{Wyrzykowski2011a}
{Wyrzykowski} {\L}.,  et~al., 2011a, \mn@doi [\mnras]
  {10.1111/j.1365-2966.2010.18150.x}, \href
  {http://adsabs.harvard.edu/abs/2011MNRAS.413..493W} {413, 493}

\bibitem[\protect\citeauthoryear{{Wyrzykowski} et~al.,}{{Wyrzykowski}
  et~al.}{2011b}]{Wyrzykowski2011b}
{Wyrzykowski} L.,  et~al., 2011b, \mn@doi [\mnras]
  {10.1111/j.1365-2966.2011.19243.x}, \href
  {http://adsabs.harvard.edu/abs/2011MNRAS.416.2949W} {416, 2949}

\bibitem[\protect\citeauthoryear{{Yang}, {Feng}, {Huang}, {Chen}, {Lu}  \&
  {Zong}}{{Yang} et~al.}{2011a}]{Yang11c}
{Yang} Y.,  {Feng} L.,  {Huang} X.,  {Chen} X.,  {Lu} T.,   {Zong} H.,  2011a,
  \mn@doi [\jcap] {10.1088/1475-7516/2011/12/020}, \href
  {http://adsabs.harvard.edu/abs/2011JCAP...12..020Y} {12, 20}

\bibitem[\protect\citeauthoryear{{Yang}, {Feng}, {Huang}, {Chen}, {Lu}  \&
  {Zong}}{{Yang} et~al.}{2011b}]{Yang12}
{Yang} Y.,  {Feng} L.,  {Huang} X.,  {Chen} X.,  {Lu} T.,   {Zong} H.,  2011b,
  \mn@doi [\jcap] {10.1088/1475-7516/2011/12/020}, \href
  {http://adsabs.harvard.edu/abs/2011JCAP...12..020Y} {12, 20}

\bibitem[\protect\citeauthoryear{{Yang}, {Huang}, {Chen}  \& {Zong}}{{Yang}
  et~al.}{2011c}]{Yang11a}
{Yang} Y.,  {Huang} X.,  {Chen} X.,   {Zong} H.,  2011c, \mn@doi [\prd]
  {10.1103/PhysRevD.84.043506}, \href
  {http://adsabs.harvard.edu/abs/2011PhRvD..84d3506Y} {84, 043506}

\bibitem[\protect\citeauthoryear{{Yang}, {Yang}, {Huang}, {Chen}, {Lu}  \&
  {Zong}}{{Yang} et~al.}{2013a}]{Yang13a}
{Yang} Y.,  {Yang} G.,  {Huang} X.,  {Chen} X.,  {Lu} T.,   {Zong} H.,  2013a,
  \mn@doi [\prd] {10.1103/PhysRevD.87.083519}, \href
  {http://adsabs.harvard.edu/abs/2013PhRvD..87h3519Y} {87, 083519}

\bibitem[\protect\citeauthoryear{{Yang}, {Yang}  \& {Zong}}{{Yang}
  et~al.}{2013b}]{Yang13c}
{Yang} Y.,  {Yang} G.,   {Zong} H.,  2013b, \mn@doi [\prd]
  {10.1103/PhysRevD.87.103525}, \href
  {http://adsabs.harvard.edu/abs/2013PhRvD..87j3525Y} {87, 103525}

\bibitem[\protect\citeauthoryear{{Yang}, {Yang}  \& {Zong}}{{Yang}
  et~al.}{2013c}]{Yang13b}
{Yang} Y.-P.,  {Yang} G.-L.,   {Zong} H.-S.,  2013c, \mn@doi [Eur.\ Phys.\
  Lett.] {10.1209/0295-5075/101/69001}, \href
  {http://adsabs.harvard.edu/abs/2013EL....10169001Y} {101, 69001}

\bibitem[\protect\citeauthoryear{{You} et~al.,}{{You} et~al.}{2007}]{You2007}
{You} X.~P.,  et~al., 2007, \mn@doi [\mnras]
  {10.1111/j.1365-2966.2007.11617.x}, \href
  {http://adsabs.harvard.edu/abs/2007MNRAS.378..493Y} {378, 493}

\bibitem[\protect\citeauthoryear{{Zackrisson} et~al.,}{{Zackrisson}
  et~al.}{2013}]{Zackrisson12}
{Zackrisson} E.,  et~al., 2013, \mn@doi [\mnras] {10.1093/mnras/stt303}, \href
  {http://adsabs.harvard.edu/abs/2013MNRAS.431.2172Z} {431, 2172}

\bibitem[\protect\citeauthoryear{{Zhang}}{{Zhang}}{2011}]{Zhang11}
{Zhang} D.,  2011, \mn@doi [\mnras] {10.1111/j.1365-2966.2011.19602.x}, \href
  {http://adsabs.harvard.edu/abs/2011MNRAS.418.1850Z} {418, 1850}

\bibitem[\protect\citeauthoryear{{Zheng}, {Yang}, {Li}  \& {Zong}}{{Zheng}
  et~al.}{2014}]{Zheng14}
{Zheng} Y.-L.,  {Yang} Y.-P.,  {Li} M.-Z.,   {Zong} H.-S.,  2014, \mn@doi
  [Res.\ in A\&A] {10.1088/1674-4527/14/10/001}, \href
  {http://adsabs.harvard.edu/abs/2014RAA....14.1215Z} {14, 1215}

\makeatother
\end{thebibliography}
\bibliographystyle{mnras}
% Don't change these lines
\bsp    % typesetting comment
\label{lastpage}


\section*{Supplementary material}

# Erratum: Investigating dark matter substructure with pulsar timing I. Constraints on ultracompact minihalos

Hamish A. Clark[1]⋆, Geraint F. Lewis[1], Pat Scott[2]
[1]*Sydney Institute for Astronomy, School of Physics A28, The University of Sydney, NSW 2006, Australia*
[2]*Department of Physics, Imperial College London, Blackett Laboratory, Prince Consort Road, London SW7 2AZ, UK*

The paper 'Investigating dark matter substructure with pulsar timing: I. Constraints on ultracompact minihalos' was published in MNRAS, 456, p. 1394 (Clark et al. 2016). Due to an error in the interpretation of the timing noise, the statistical limits on UCMH abundance published in Fig. 5 are no longer valid. This error stems from the use of the first derivative of the gravitational time delay as a contribution to a pulsar's period derivative. This would actually contribute instead to its measured period.

Accounting for this, we have compared the expected strength of both the first and second derivative timing noise ($P$-noise and $\dot{P}$-noise) to ATNF data. Unfortunately, the difference between the period derivative and period of pulsars is on average around 15 orders of magnitude, and as such we have found that neither is sufficient to rule out even the scenario that the Milky Way's dark matter consists entirely of UCMHs ($f = 1$). Therefore, this method (as described in Section 2.3) does not allow any meaningful limits to be placed on the abundance of UCMHs within the Milky Way.

Despite this, the projected limits on $f$ from non-detection of individual UCMH timing signals in millisecond pulsar data (Fig. 3) are unaffected, and thus remain valid. At its strongest, the individual-detection method is able to provide a projected constraint of $f \lesssim 0.01$ at a mass of $10^{-3} M_\odot$ – assuming a detection threshold of 1 ns. For wavenumbers $k \gtrsim 10^4$ Mpc$^{-1}$, these are the strongest projected limits on UCMH abundance to date that do not rely upon the assumption that dark matter annihilates. This method allows us to probe down to a mass that is lower than attainable by other gravitational means – a quality which assists significantly in providing the strongest limits on primordial power. Comparatively, the projected limits of Zackrisson et al. (2013) and Li et al. (2012) are only able to constrain (at best) $f \lesssim 0.1$ and $f \lesssim 0.01$ for halo masses of approximately $10^6 M_\odot$ and $10^{-1} M_\odot$, respectively.

## REFERENCES

Clark H. A., Lewis G. F., Scott P., 2016, MNRAS, 456, 1394
Li F., Erickcek A. L., Law N. M., 2012, Phys. Rev. D, 86, 043519
Zackrisson E., et al., 2013, MNRAS, 431, 2172

This paper has been typeset from a T<sub>E</sub>X/LAT<sub>E</sub>X file prepared by the author.

⋆ E-mail: hamish.clark@sydney.edu.au (HAC)


\end{document}